\documentclass{aastex6}
\usepackage{eqnarray,amsmath}

\begin{document}

%% LaTeX will automatically break titles if they run longer than
%% one line. However, you may use \\ to force a line break if
%% you desire.

%\title{Concavity of the hamiltonian of a quantized photon propagating in the constant magnetic field}
\title{The Anomalous Magnetic Moment of a photon propagating in a magnetic field}
%% Use \author, \affil, plus the \and command to format author and affiliation 
%% information.  If done correctly the peer review system will be able to
%% automatically put the author and affiliation information from the manuscript
%% and save the corresponding author the trouble of entering it by hand.
%%
%% The \affil should be used to document primary affiliations and the
%% \altaffil should be used for secondary affiliations, titles, or email.

%% Authors with the same affiliation can be grouped in a single
%% \author and \affil call.
\author{Julian W. Mielniczuk\altaffilmark{1}, S. R. Valluri \altaffilmark{2,3}, Darrell Lamm \altaffilmark{4} and Sayantan Auddy  \altaffilmark{2} }

\altaffiltext{1}{82 Brick Kiln Road Apt 1-301, Chelmsford, M -01824, USA}~\\
\altaffiltext{2}{Department of Physics and Astronomy, University of Western Ontario, London, N6A 3K7, Canada}
\altaffiltext{3}{King's University College, University of Western Ontario, London, N6A 3K7, Canada}
\altaffiltext{4}{Georgia Institute of Technology, Atlanta, GA 30332, USA}

%\author{Butler Burton\altaffilmark{3}}
%\affil{National Radio Astronomy Observatory}

%\author{Amy Hendrickson}
%\affil{TeXnology Inc}

%\author{Julie Steffen\altaffilmark{4}}
%\affil{American Astronomical Society \\
%2000 Florida Ave., NW, Suite 300 \\
%Washington, DC 20009-1231, USA}

%% Use the \and command so offset the last author.
%\and

%\author{Jeff Lewandowski\altaffilmark{5}}
%\affil{IOP Publishing, Washington, DC 20005}

%% Notice that each of these authors has alternate affiliations, which
%% are identified by the \altaffilmark after each name.  Specify alternate
%% affiliation information with \altaffiltext, with one command per each
%% affiliation.

%\altaffiltext{1}{AAS Journals Data Scientist}
%\altaffiltext{2}{greg.schwarz@aas.org}
%\altaffiltext{3}{AAS Journals Associate Editor-in-Chief}
%\altaffiltext{4}{AAS Director of Publishing}
%\altaffiltext{5}{IOP Senior Publisher for the AAS Journals}

%% Mark off the abstract in the ``abstract'' environment. 
\begin{abstract}
We analyze the spectrum of the Hamiltonian of a photon propagating in a strong magnetic field $B\sim B_{\rm{cr}}$, where $B_{\rm cr}= \frac{m^2}{e} \simeq 4.4 \times 10^{13}$ Gauss is the  Schwinger critical field .
We show that the expected value of the Hamiltonian of a quantized photon for a perpendicular mode is a concave function of the magnetic field $B$. We show by a partially analytic and numerical method that the anomalous magnetic moment of a photon in the one loop approximation is a non - decreasing function of the magnetic field $B$ in the range $0\leq B \leq 30 \, B_{\rm cr}$ We provide a numerical representation of the expression for the anomalous magnetic moment in terms of special functions. We find that the anomalous magnetic moment $\mu_\gamma$ of a photon for $B=30\, B_{\rm cr }$ is $8/3$ of the anomalous magnetic moment of a photon for $B = 1/2 ~ B_{\rm cr}$.

\end{abstract}

%% Keywords should appear after the \end{abstract} command. 
%% See the online documentation for the full list of available subject
%% keywords and the rules for their use.
%\keywords{editorials, notices --- 
%miscellaneous --- catalogs --- surveys}
%% From the front matter, we move on to the body of the paper.
%% Sections are demarcated by \section and \subsection, respectively.
%% Observe the use of the LaTeX \label
%% command after the \subsection to give a symbolic KEY to the
%% subsection for cross-referencing in a \ref command.
%% You can use LaTeX's \ref and \label commands to keep track of
%% cross-references to sections, equations, tables, and figures.
%% That way, if you change the order of any elements, LaTeX will
%% automatically renumber them.
%% We recommend that authors also use the natbib \citep
%% and \citet commands to identify citations.  The citations are
%% tied to the reference list via symbolic KEYs. The KEY corresponds
%% to the KEY in the \bibitem in the reference list below. 

\section{Introduction} \label{sec:intro}
The nonlinearity  of Maxwell's wonderful equations continues to present and challenge us with a variety of interesting phenomenon. The effective interaction that results due to the corrections from the virtual excitations of the charged quantum fields, such as electron $e^{-}$ and positron $e^{+}$, leads to well known interesting effects \citep{dit00}.  More recently, other interesting aspects of the quantum vacuum have been explored by \cite{sha11,cha12,alt08} to name but a few. In the case of electromagnetic fields that vary slowly with respect to the Compton wavelength, i.e. frequencies much less than the pair creation threshold, the one loop quantum electrodynamic effective Heisenberg-Euler Lagrangian  (HEL), \cite{mk79,sha11,cha12,dun04} describes the dominant physical effects. The HEL is known to all orders in electromagnetic fields. It is well known that electrons acquire an anomalous magnetic moment due to the radiative corrections in quantum electrodynamics (QED) with the $e^{-} - e^{+}$ pairs and virtual photons in the background \citep{sch51}. It is also of great fundamental interest that there is an anomalous photon magnetic moment $\mu_\gamma$ due to the interaction with the external magnetic field in the environment of the virtual $e^{-} - e^{+}$ quanta of the vacuum. The last couple of decades has seen a resurgence of interest in quantum vacuum physics \citep{gie07,bar95,mie88,hey97a,hey97b,hey97c,dun09}.The promise of high intensity experimental facilities ($\sim 10^{15}$ W) has stimulated immense interest and enthusiasm to investigate the nonlinear quantum vacuum in practical optical experiments \citep{mar06,dun09,val13,val14}. The Polarization of the Vacuum with Laser (PVLAS) experiment aims to measure the birefringence of the external magnetic field in the vacuum \citep{zav08,bre08,can08}.

In section 2, we outline and discuss the analytic calculations on the anomalous  magnetic moment of the photon. We present and discuss the results. In section 3 we briefly outline the mathematical expression for the photon center of mass and the expression for the group velocity. Section 4 presents the conclusions. The supplementary mathematical details are provided in appendices A, B and C.

\section{Anomalous Magnetic Moment of a Photon}
%\latex\ \footnote{\url{http://www.latex-project.org/}} is a document markup
\cite{cha10,cha12} and \cite{roj06,roj07} have discussed the notion of the anomalous magnetic moment of a photon . The photon anomalous magnetic moment and its paramagnetic properties that have been studied by \cite{roj14,roj06,roj07} have provided values of $\mu_\gamma$ in the two extreme limits of $B \ll B_{\rm cr}$ and $B \gg B_{\rm cr}$. The purpose of this paper is to provide numerical values and an analytic formula for the range $B \sim B_{\rm cr}$. Our results are applicable in the range $0\leq B \leq 30 \, B_{\rm cr}$.

At one-loop order, the Heisenberg Euler effective Lagrangian in constant external electromagnetic fields \citep{hei36,kar15}, describing the effective nonlinear interactions between the electromagnetic fields mediated by electron-positron fluctuations in the vacuum, can be represented concisely in terms of the following proper time integral \citep{sch51}.
\begin{equation}\label{Lagrangian}
\mathcal{L} = \frac{\alpha}{2\pi}\int_0^{\infty}\frac{ds}{s} e^{{-i\frac{m^2}{e}s}} \left[ab\coth(as)\cot(bs)- \frac{a^2-b^2}{3}-\frac{1}{s^2}\right]
\end{equation}
with the prescription $m^2 \rightarrow m^2-\i0^{+}$, and the proper time integration contour assumed to lie slightly below the real positive $s$ axis. Here, m is the electron mass, $e$ is the elementary charge, $\alpha=\frac{e^2}{4\pi}$ is the fine structure constant, and $a=(\sqrt{\mathcal{F}^2+\mathcal{G}^2}-\mathcal{F})^{1/2}$ and $b=(\sqrt{\mathcal{F}^2+\mathcal{G}^2}+\mathcal{F})^{1/2}$ are the secular invariants made up of the gauge and Lorentz invariants of the electromagnetic field: $\mathcal{F}= \frac{1}{4}F_{\mu \nu}F^{\mu \nu}=\frac{1}{2}(\textbf{B}^2-\textbf{E}^2)$ and $\mathcal{G}=\frac{1}{4}F^*_{\mu \nu}F^{\mu \nu}= -\textbf{E}\cdot \textbf{B}$, with $^{*}F^{\mu\nu}= \frac{1}{2}\epsilon^{\mu \nu \alpha \beta}F_{\alpha \beta}$ denoting the dual field strength tensor; $\epsilon^{\mu \nu \alpha \beta}$ is the totally antisymmetric tensor, fulfilling $\epsilon^{0123}=1$. Our metric convention is $ g_{\mu \nu} = diag(-1,+1,+1,+1)$, and we use the units where $c =\hbar =1$. To keep notations compact we moreover employ the short hand notations $\int_{x} \equiv \int d^4 x$ and $\int_k\equiv \int \frac{d^4 k}{(2\pi)^4)}$ for the integration over the position and the momentum space, respectively.
%The effective Lagrangian(\citep{•}) is a gauge and Lorentz invariant quantity.  Also clearly, for inhomogeneous background fields additional gauge and Lorentz building blocks become available. For slowly varying fields the deviations from the constant field limit can be accounted for with a derivative term $\sim \partial_{\alpha}F^{\mu \nu}$.
\par The seminal paper of \cite{sch51} on gauge invariance and vacuum polarization has used the proper time parameter formulation to the solution of the equation of motion of a particle. Thereby, the effective Lagrangian \citep{kar15} is finite, gauge and Lorentz invariant. The derivative expansion of the one loop effective Lagrangian in QED has been studied by \cite{gus96}. Their non-perturbative term is that derived by Schwinger but the second term in their expansion shows explicitly the two derivatives of $F_{\mu \nu}$ that account the case where the fields are slowly or fast varying. We do not consider them now in the assumption of the constant field approximation but the effects of additional terms to the Schwinger Lagrangian warrant a more detailed analysis in a further study. If the typical frequency/momentum scale of the variation of the homogeneous background field is $\nu$, derivatives effectively translate into multiplications with $\nu$ to be rendered dimensionless by the  electron mass $m$. Thus, Equation \ref{Lagrangian} is also applicable for slowly varying inhomogeneous fields fulfilling $\frac{\nu}{m} \ll 1$, or in other words for inhomogeneities whose typical spatial (temporal) scales of variation are much larger than the Compton wavelength (time) $\sim \frac{1}{m}$ of the virtual charged particle. The electron Compton wavelength is $\lambda_{c}=3.86 \times 10^{-13}$ m and the Compton time is $\tau_{c} = 1.29 \times 10^{-21}$ s. In turn, many electromagnetic fields available  in the laboratory, e.g., the electromagnetic field pulses generated by optical high intensity lasers, \citep{dun09} featuring wavelengths of $\mathcal{O}(\mu m)$ and pulse durations of $\mathcal{O}(fs)$, are compatible with this requirement. 
\par The effective Lagrangian is a scalar quantity, and the scalar quantities made up of combinations of $F^{\mu \nu}$, $^* F^{\mu \nu}$, and the derivatives thereof involve an even number of derivatives. Hence, when employing the constant fields results \citep{kar15} for the slowly varying inhomogeneous field, the derivations from the corresponding exact results are of $\mathcal{O}((\frac{\nu}{m})^2)$. In the absence of an external electric field the partial derivatives of the effective action in the one loop approximation are \citep{lun09,lun10}

\begin{subequations}\label{second:main}
\begin{equation}
\gamma_{\mathcal{F}}= \frac{\partial \mathcal{L}}{\partial {\mathcal{F}}}, \gamma_{\mathcal{F}\mathcal{F}}=\frac{\partial^2 \mathcal{L}}{\partial \mathcal{F}^2},
\gamma_{\mathcal{G}\mathcal{G}}=\frac{\partial^2 \mathcal{L}}{\partial \mathcal{G}^2}. 
\tag{2}
\end{equation}

\begin{flushleft}
Expressions such as $\gamma_{\mathcal{G}}$, $\gamma_{\mathcal{F}\mathcal{G}}$ are zero for zero electric field. Further,
\end{flushleft}

\begin{equation}
\gamma_\mathcal{F}= -1-\frac{\alpha}{2\pi}[\frac{1}{3}+2h^2-8\zeta^{\prime}(-1,h)+4h\ln\Gamma(h)-2h\ln h+\frac{2}{3}\ln h-2h\ln2\pi]\label{second:a}
\end{equation}

\begin{equation}
\gamma_{\mathcal{F}\mathcal{F}}=\frac{\alpha}{2\pi{B^{2}}}[\frac{2}{3}+4h^2\psi(1+h)-2h-4h^2-4h\ln\Gamma(h)+2h\ln2\pi-2h\ln h]\label{second:b}
\end{equation}

\begin{equation}
\gamma_{\mathcal{G}\mathcal{G}} = \frac{\alpha}{2 \pi B^{2}}\left[- \frac{1}{3}-\frac{2}{3}\psi (1+h) - 2h^{2}+(3h)^{-1}+8 \zeta^{\prime}(-1,h)-4h\ln \Gamma(h)+2h\ln(2\pi)+2h\ln h \right]\label{second:c}
\end{equation}  

\end{subequations}

where $\psi $ is the digamma function, $\Gamma$ is the gamma function, and $h=\frac{1}{2}\frac{B_{\rm c}}{B}$.
\begin{equation}
 \gamma_{\mathcal{G}\mathcal{G}} =\frac{\partial^2 \mathcal{L} }{\partial \mathcal{G}^2} \arrowvert_{\mathcal{F}=\frac{1}{2}B^2}^{\mathcal{G}=0}
\end{equation}

Also
\begin{equation}
\zeta^{\prime} (s,h)= \partial_s \zeta(s,h),
\end{equation}
where $\zeta(s,h)$ is the Hurwitz zeta function, for $s=-1$ given by \cite{ada04} and $h>>1$ \citep{dit79}

\begin{equation}
\zeta^{\prime}(-1,h)\cong \frac{1}{12}-\frac{h^2}{4}+\frac{\ln h}{2}(h^2-h+\frac{1}{6})+ \int^{\infty}_{0} \frac{e^{-hx}}{x^2}(\frac{1}{1-e^{-x}}-\frac{1}{x}-\frac{1}{2}-\frac{x}{12})dx, \,\,\, Re(h)>0
\end{equation}

\begin{equation}
\zeta^{\prime}(-1,h)\cong \frac{1}{12}-\frac{h^2}{4}+\frac{\ln h}{2}(B_2(h))+\frac{1}{720}\frac{1}{h^2}.
\end{equation}where $B_2(h)=h^2-h+\frac{1}{6}$ is the second Bernoulli polynomial \citep{olv10}. The integral above is convergent \citep{ada04}.

The refractive indices for perpendicular and parallel polarized photons are of particular interest in this context. It is worth noting that 
\begin{equation}
\frac{4\pi}{\alpha}(n_{\perp} -1)=\frac{2\pi{B}^2}{\alpha}\gamma_{\mathcal{G}\mathcal{G}},
\end{equation}
where $\gamma_{\mathcal{G}\mathcal{G}}$ has been defined in Equation {\ref{second:c}}.
For the weak field case, $n_{\perp}$ is given by the expression \citep{hey97a,hey97b,hey97c}.
 $\xi = \frac{B}{B_{cr}} = \frac{1}{2h}$ and $\xi<1$
\begin{equation}
n_{\perp} = 1+ \frac{\alpha}{4\pi}\sin^{2}\theta \left[ \frac{14}{45}\xi^{2} - \frac{1}{3} \sum_{j=2}^{\infty} \frac{2^{2j}(6B_{2(j+1)}-(2j+1)B_{2j})}{j(2j+1)} \xi^{2j}\right] + \mathcal{O}\left[\left(\frac{\alpha}{2\pi}\right)^2\right]
\end{equation}
where $\xi = \frac{B}{B_{\rm cr}}$, $\alpha$ is the fine structure constant, $B_{2j}$ and $B_{2j+1} $ are the Bernoulli numbers.
In the strong-field limit $(\xi > 0.5)$, we obtain

\begin{eqnarray}
n_{\perp} = 1+\frac{\alpha}{4\pi}\sin^{2}\theta [ \frac{2}{3}\xi - \left(8 \ln A -\frac{1}{3} - \frac{2}{3}\gamma \right) - \left(\ln\pi +\frac{1}{18} \pi^{2}-2-\ln\xi\right) \xi^{-1} \\ - \left(-\frac{1}{2} - \frac{1}{6}\zeta(3)\right)\xi^{-2}-\sum_{j=3}^{\infty}\frac{(-1)^{j-1}}{2^{j-2}}\left[\frac{j-2}{j(j-1)}\zeta(j-1)+\frac{1}{6}\zeta(j+1)\xi^{-j} \right] + \mathcal{O}\left[\left(\frac{\alpha}{2\pi}\right)^2\right]
\end{eqnarray}
For parallel polarizations, the refractive index is given by \cite{tsa75}

\begin{equation}
n_{\parallel}=1+\frac{\alpha}{4\pi}[-\frac{1}{3}-\frac{2}{3}\psi(1+h)+8\zeta^{\prime}(-1,h)-2h^2+\frac{1}{3h}-4h \ln\Gamma(h)+2h\ln(2\pi)+2h\ln h].
\end{equation}

\begin{flushleft}

which is valid for all $B \leq \frac{\pi}{\alpha} B_{cr}$.
\end{flushleft}
\begin{eqnarray}
\Delta{n}_{\perp,\parallel} = n_{\perp}-n_{\parallel}=\frac{\alpha}{4\pi}[ \frac{1}{3h} - (8 \ln A -\frac{1}{3} - \frac{2}{3}\gamma) -2h (\ln\pi +\frac{\pi^2}{18}-2+\ln2h) \nonumber \\ +2h^2+\frac{2}{3}\zeta(3)h^2 -\sum_{j=3}^{\infty}\frac{(-1)^{j-1}}{2^{j-2}} +\lbrace (j-2)+\frac{\zeta(j+1)}{6}+\frac{h^j}{2j} \rbrace + \frac{2}{3}\ln(1+h)-\frac{1}{3}\frac{1}{1+h}-\frac{2}{3}-\frac{2}{3}\ln h-\frac{22}{48}\frac{1}{h^2}]\nonumber \\
\end{eqnarray}

Here $\zeta$ is the Riemann zeta function, $\zeta(3)\cong 1.202$, $\theta$ is the angle between the magnetic field $\textbf{B}$ and the vector $\textbf{k}$, $\gamma \cong 0.577$ which is the Euler-Mascheroni constant and $A \cong 1.28242712...$ is the Glaisher-Kinkelin constant \citep{olv10}.

An important physical variable is the Faraday rotation angle $\chi$ as $\chi=k(n_{\perp}-n_{\parallel})l$, where $k$ is the magnitude of the photon wave vector and $l$ can be viewed as the path distance of the photon in the magnetic field. The Faraday rotation can, in principle, be observable for appreciable values of k and $l$ . \par
We will analyze the properties of a photon propagating in a strong magnetic field $\textbf{B}$. The Hamiltonian of a photon is given by \citep{bia12,bia14}

\begin{equation}\label{12}
\hat{H}(B) = \sum_{\lambda} \int d^3k\, \hbar \omega_{k} a^{\dagger}_{\lambda}(\textbf{k})a_{\lambda}(\textbf{k})
\end{equation} 

\begin{flushleft}
where the creation and the annihilation operator satisfy the commutation rule
\end{flushleft}

\begin{equation}
[a^{\dagger}_{\lambda}(\textbf{k}), a_{\lambda^{\prime}}(\textbf{k}^{\prime})]= \delta_{{\lambda},{\lambda^{\prime}}} \delta(\textbf{k} - \textbf{k}^{\prime})
\end{equation}

\begin{flushleft}
 From the linearity in the term proportional to $\mu_{\gamma}$ of the Hamiltonian \citep{cha12,roj14}.
\end{flushleft}

\begin{equation}\label{14}
\mu_{\gamma} = - \frac{d\langle \hat{H}(\textbf{B})\rangle}{d B}
\end{equation} 
where $\mu_\gamma$ denotes the magnetic moment and $\langle \rangle$ denotes the quantum expectation value for a perpendicularly polarized photon. We have
\begin{eqnarray}
\omega_{\parallel} = \frac{| \textbf{k} |}{n_{\parallel}} \\
\omega_{\perp} =  \frac{| \textbf{k} |}{n_{\perp} }
\end{eqnarray}
 $\omega_{\parallel}$ and $\omega_{\perp}$ are the photon frequencies in the parallel and the perpendicular modes and the corresponding indices of refraction are

\begin{equation}
n_{\parallel}=\frac{1}{\sqrt{1-\kappa_{s}\sin^2 \theta} } 
\end{equation}

\begin{equation}\label{16}
n_{\perp}=\sqrt{\frac{1+\kappa_p}{1+\kappa_{s}\cos^2 \theta} } 
\end{equation}
where $\kappa_{s}$ and $\kappa_{p}$ are given below by equations (\ref{kappas}) and (\ref{kappap}).
For $\theta = \frac{\pi}{2}$ we have 
\begin{equation}
n_{\perp} = \sqrt{1+\kappa_{p}}
\end{equation}
\begin{equation}\label{kappas}
\kappa_s = \gamma_{\mathcal{F}\mathcal{F}}\frac{B^2}{\gamma_{s}}.
\end{equation}
\begin{equation}
\gamma_s= 1-\gamma_{\mathcal{F}}
\end{equation}
Using the binomial expansion, $n_{\parallel}$ can be approximately written as
\begin{equation}
n_{\parallel}=1-\frac{1}{2}B^{2}\gamma_{\mathcal{F}\mathcal{F}}.
\end{equation}

\begin{flushleft}
we will approximate \end{flushleft}
\begin{equation}
\frac{1}{n_{\perp}} = \frac{1}{\sqrt{1+\kappa_p}} \cong 1 - \frac{1}{2}\kappa_p
\end{equation}
and $\gamma_s \cong 1$.
So  
\begin{equation}\label{24}
\langle H(B)\rangle \cong \ \langle H(0)\rangle - \frac{1}{2}B^{2}\gamma_{\mathcal{G}\mathcal{G}}
\end{equation}
\begin{equation}\label{kappap}
\kappa_{p} =\gamma_{\mathcal{G} \mathcal{G}} B^2/ \gamma_s
\end{equation}
Following \cite{bia12,bia14} we will call the mode perpendicular if the magnetic field of the photon is in the plane formed by the vectors $\textbf{B}$ and $\textbf{k}$ where $\textbf{k}$ is the wave vector of the photon. In the approximation
\begin{equation}
\frac{1}{\sqrt{1+\kappa_p}} \cong 1 -\frac{1}{2}\kappa_p
\end{equation} 
and $\gamma_s \cong 1 $ we will confine ourselves to the range $0\leq B \leq \, B_{\rm cr}$. The radiative corrections come into visible play for $B \geq 430 B_{\rm cr}$. As an aside, it is interesting to note that although the equation of motion of a neutrino in an external magnetic field is effectively altered \citep{mk91}, the radiative correction effects on a neutrino beam by a strong magnetic field is $2\times10^{10} T$ have been found to be extremely small. 

We define 
\begin{equation}
\gamma_{\mathcal{F}\mathcal{F}}= \frac{\partial^2 \mathcal{L} }{\partial \mathcal{F}^2} \arrowvert_{\mathcal{F}=\frac{1}{2} B^2 }^{\mathcal{G}=0}
\end{equation}

\begin{figure}[h]
	\label{fig1}
\centerline{\includegraphics[height=6cm]{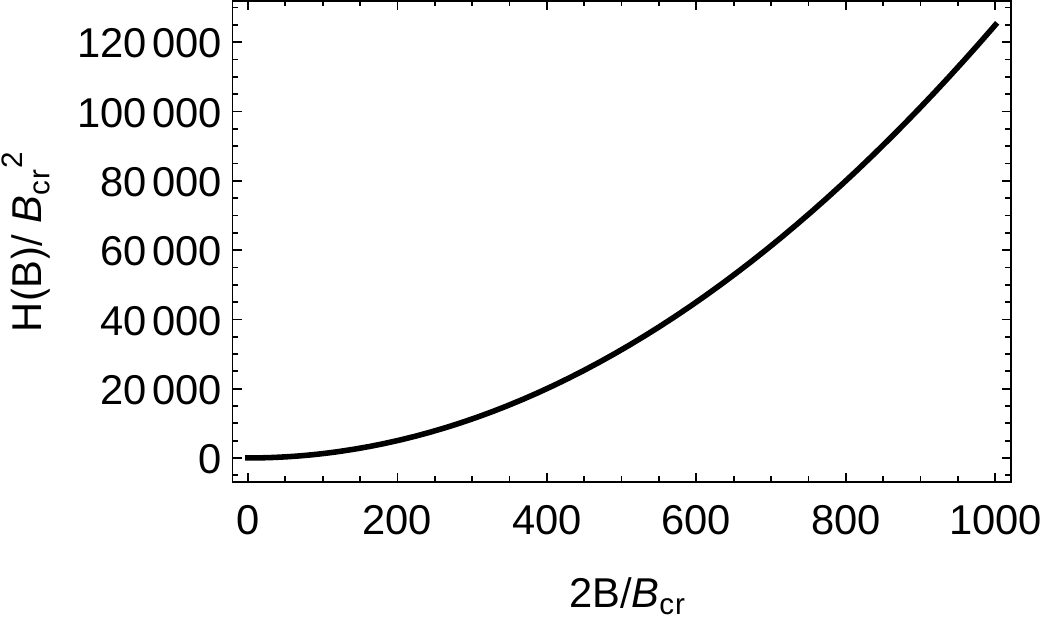}}
\caption{Shows the expression of the Hamiltonian given in \ref{24} as a function of the magnetic field B. The vertical and the horizontal axis are normalized in units of $B_{cr}^2$ and $B_{cr}$, respectively.}
\end{figure}

\begin{flushleft}
Fig.(1) illustrates that $\langle H(B)\rangle $ is a convex function of the magnetic field $\textbf{B}$. 
The numerical results of Fig. (1) were obtained by \textit{Mathematica 10.} Fig.(1) illustrates that $\langle H (B) \rangle$ is a increasing function of the magnetic field. 
\end{flushleft}

From equations \ref{14} and \ref{24} the photon magnetic moment of a perpendicularly polarized photon for $B \leq 30 B_{\rm cr}$ and the fact that $\mu_\gamma(0)=0$, is given by
\begin{equation}\label{2.1}
\mu_\gamma(B) = \frac{\alpha}{4\pi} \left\lbrace\frac{2}{3}+\frac{1}{B^3}\left[\frac{1}{3}B \psi^{\prime} \left(1+\frac{1}{2B}\right)+ \psi\left(\frac{1}{2B}\right)-2B\ln \Gamma \left(\frac{1}{2B}\right)+ B\ln(2\pi)+ B +B \ln(2B)-1) \right] \right\rbrace \left(\frac{|\textbf{k}|}{m}\right)\sin^2 \theta
\end{equation}
where $\psi$ is the digamma function, $\Gamma$ is the Euler gamma function. From equation (\ref{2.1}), one observes that the photon magnetic moment contributes to both the external field strength as well as the photon energy through its momentum.

\begin{flushleft} For $B > \frac{1}{2}B_{\rm cr}$ we can approximate \end{flushleft}
\begin{equation}\label{2.3}
\mu_{\gamma}(B) \cong \frac{\alpha}{4 \pi} \left[\frac{2}{3}+ \left(\ln(\pi)+\frac{\pi^2}{18}-1\right) B^{-2} - \frac{\ln B}{B^2}\right] \left(\frac{|\textbf{k}|}{m}\sin^2 \theta \right)
\end{equation}
where  $\textbf{k}$ is the photon wave vector.
 
It is interesting to note that rearrangement of equation (\ref{2.3}) with the terms involving $B$ and setting $c_1= \rm{ln} \pi +\frac{\pi^2}{18} -1$ gives the following equivalent expression of $\rm{ln} B$:
\begin{equation}
2(c_1-\rm{ln} B)=W_j\left( \left\lbrace -\frac{8 \pi}{\alpha}\frac{m}{|k|sin^2 \theta} \mu_{\gamma}(B)+\frac{4}{3} \right\rbrace \ {e^{2^{c_1}}}\right),
\end{equation}
where $W_j$ denotes the $j$th branch of the multivalued inverse function known as the Lambert W function \citep{val00}. The Lambert W function is defined such that \citep{cor96}

\begin{equation}
W(z)e^{W(z)}=z ,
\end{equation}
where $z$ can be a complex variable. The utility of this function in QED is an aspect that warrants study, although it has found many remarkable applications in a multitude of diverse fields \citep{cor96,vall09,rob16}

\begin{flushleft}For $0\leq B \leq 0.44 B_{\rm cr}$\end{flushleft}
\begin{equation}\label{2.4}
\mu_\gamma(B) \cong \frac{\alpha}{4 \pi} \frac{28}{45} \left(B-\frac{52}{49}B^{3}\right) \left(\frac{|\textbf{k}|}{m}\sin^2 \theta\right).
 \end{equation}

\begin{flushleft}
For a perpendicularly polarized photon, we note that equation (\ref{2.4}) can be replaced by the inequality 
\end{flushleft}

\begin{equation}
\mu_\gamma(B) \geq \frac{\alpha}{4 \pi} \frac{28}{45} \left(B-\frac{52}{49}B^{3}\right) \left(\frac{|\textbf{k}|}{m}\sin^2 \theta\right)
\end{equation}
We restrict  equation (\ref{2.1}) to $0\leq B \leq 30 B_{\rm cr}$.
Using equation (\ref{2.3}) we obtain that $\mu_\gamma (B=30 B_{\rm cr})$ is only $ 3 \%$   smaller than the asymptotic value $\alpha/(3\pi) $ of the Bohr magneton. It is approximately $10^{-3} $ of the Bohr magneton for $|\textbf{k}|\sim m$.
$\mu_\gamma(B)$ grows from the value of 

\begin{equation}
\frac{\alpha}{4\pi}\frac{28}{45}\frac{3}{4}\frac{1}{2}\frac{|\textbf{k}|}{m}\sin^2 \theta
\end{equation}
for $B=\frac{1}{2}B_{\rm cr}$ to the value very close to 
\begin{equation}
\frac{\alpha}{4\pi	}\frac{2}{3}\frac{|\textbf{k}|}{m}\sin^2 \theta,
\end{equation}
for $B=30 B_{cr}$ so the growth is only by a factor of $\approx 3 $ . Equation {\ref{2.3}} is the generalization of equation (157) of \cite{cha12}, who state that 
\begin{equation}
\mu_\gamma(B)\sim \frac{\alpha}{3\pi}\left(\frac{1}{2}\frac{e}{m}\right),
\end{equation}
for large values of the magnetic field $\textbf{B}$. This suggest that the one loop approximation provides a good estimate of $\mu_{\gamma}$ in the low frequency case. Here $e$ denotes the electron charge and m is the corresponding mass. At low and high photon frequency \cite{cha06} have shown that the photon magnetic moment shows a paramagnetic behavior as is also true for the vacuum embedded in a strong external magnetic fields \citep{mie88}. Our equation \ref{2.4} is similar to equation {\ref{16}} of  \cite{roj14} except that our numerical factor $\frac{28}{45}$ is twice bigger than their corresponding factor $\frac{14}{45}$ . Formally our equation  is applicable only when
\begin{equation}\label{2.8}
\frac{|\textbf{k}|}{m} \ll 1
\end{equation}
Equations (\ref{2.1}) and (\ref{2.3}) are the main results of our paper. We will show analytically in Appendices A and B that

\begin{equation}\label{2.6}
\mu_\gamma(B)> 0 \, \,\rm{for}  \, \,\, B>0.
\end{equation}
As was previously shown by \cite{roj14},
\begin{equation}\label{2.7}
\frac{d}{dB}\mu_\gamma(B) > 0 \,\,\rm{for}       \, \, \, B>0
\end{equation}
 in the two ranges $0 \leq B \leq \frac{1}{2} B_{\rm{cr}}$ and $2 \leq B$.
Equation \ref{2.7} has been checked for all positive values of $B $.
\begin{equation}
\mu_\gamma(B)=\frac{\alpha}{4\pi} \left\lbrace\frac{2}{3}+\frac{1}{B^3} \left\lbrace\left[ \frac{2}{3}B \zeta \left(2,1+\frac{1}{2B}\right)- \zeta\left(1,1+\frac{1}{2B}\right)\right\rbrace - 2B \ln \Gamma\left(\frac{1}{2B}\right) + B \left( \ln(2\pi)+1-\ln \left(\frac{1}{2B}\right) \right) -1 \right]  \right\rbrace.
\end{equation}

\begin{flushleft}where $\zeta $ is the Hurwitz zeta function. The magnetic moment can be expressed in terms of other special functions. The paramagnetic behaviour is a physical effect due to the effect of the external magnetic field on the virtual $e^{-}-e^{+}$ pairs. \end{flushleft}

\section{Photon Center of Mass}

The speed $v_{\rm cm}$ of the center of mass is proportional to $1-\frac{1}{2}B^2_{\mathcal{LGG}}$. The magnetic moment of the photon plays the leading role in determining the evolution of the photon angular momentum \citep{pry35,cha12}. The average of the corresponding center of mass location of a photon can be analyzed using the operator
\begin{equation}\label{2.9}
\hat{R}= \frac{1}{2\hat{H}}\hat{N}+\hat{N}\frac{1}{2\hat{H}}
\end{equation}

\begin{flushleft}The Hamiltonian $\hat{H}$ of \cite{haw01,haw05} is originally the Hamiltonian of a free photon. It will be replaced by our Hamiltonian $\hat{H}(B)$.\end{flushleft} For brevity we keep the same symbol,
\begin{equation}\label{2.10}
\hat{N} = \int d^3r \, \textbf{r} \, \epsilon(\textbf{r},t)
\end{equation}

\begin{flushleft}where $\hat{N}$ is the first moment of the energy distribution.\end{flushleft}
\begin{equation}
\hat{\epsilon} (\textbf{r},t) = \hat{F}^{\dagger} (\textbf{r,t}) \hat{F}(\textbf{r,t})
\end{equation}
\begin{equation}
\hat{F}(\textbf{r},t)= \frac{\hat{D}(\textbf{r},t)}{\sqrt{2\epsilon}}+i \frac{\textbf{B(\textbf{r},t)}}{\sqrt{2\mu}}
\end{equation}
\begin{equation}
x_{\rm cm} = \frac{1}{\rho^{o}}\int d^3 x \textbf{x} \Theta^{oo}
\end{equation}
and the corresponding velocity is the velocity of energy transport
\begin{equation}
v_{\rm cm2} =\frac{1}{u_2} \textbf{n} - \frac{2 \mathcal{F}\mathcal{L}\_{\mathcal{G}\mathcal{G}}}{\epsilon_{||}u_{2 \perp}}
\end{equation}

\begin{flushleft}The group velocity is less than that of the speed of light $c$, in accord with the principle of causality \citep{cha12}, with $u_{\lambda\perp} = \frac{\rho^{o}(\lambda)}{|\rho(\lambda)|}$. Here $\Theta^{oo}$ and $\rho^{o}$ are given by equations (52) and (58) of the paper by \cite{cha12}. The speed of a perpendicularly polarized photon is found to be (c=1),\end{flushleft}

\begin{equation}
v_{\perp}^2=\frac{1}{n_{\perp}^2} 
\\
\geq \frac{1}{\left(1+\frac{\alpha}{4\pi}\left(\frac{2}{3}-2h\ln h+2h\ln(2\pi)\right)\right)^2}.
\end{equation}

In the limit of ultra strong magnetic fields, the expression of $v_{\perp}^2$ when ${\theta}=\frac{\pi}{2}$, derived by \cite{hu07} is given below.

\begin{equation}
v_{\perp}^2 \simeq \frac{1-\frac{e^2}{12 \pi^2}(\ln\left(\frac{eB}{m^2}\right)-0.79)}{1-\frac{e^2}{12 \pi^2}(\ln\left(\frac{eB}{m^2}\right)-1.79)}.
\end{equation}

\section{Conclusions}
We have shown that the anomalous magnetic moment of a photon for $B=30 B_{\rm cr}$ is $8/3$
of the anomalous magnetic moment of a photon for $B=\frac{1}{2}B_{\rm cr}$. 
At low and high photon frequencies the photon magnetic moment shows a paramagnetic behavior. We find that the one loop Lagrangian is a good approximation in the range of magnetic fields considered. We have shown that the anomalous magnetic moment of a photon is a non-decreasing function of the magnetic field B for $0\leq B \leq 30 B_{\rm cr}$.\par The photon behaves like a massive pseudo vector particle under the influence of the virtual $e^{-}-e^{+}$ vacuum \citep{cha12,roj14}. Light propagation in the magnetized vacuum is analogous to the dispersion of light in an anisotropic medium. The reason for the anisotropy is due to the breaking of symmetry due to the choice of B along a preferred direction. The magnetic moment of the photon might have both astrophysical and cosmological consequences. In the presence of magnetic fields around astrophysical objects such as magnetars, magnetic lensing may be a strong observable effect.

	Photons that go by a strongly magnetized star would undergo an deflection besides the well known gravitational shift caused by the stellar mass  \citep{cha06}. The Cosmic Microwave Background (CMB) spectrum shows a substantial polarization dependent field in the vicinity of magnetars \citep{bia14}. \cite{bia14} have estimated the polarization dependent heating of the cosmic microwave background (CMB) radiation due to  strong magnetic fields. Although the large magnetic fields around the region of magnetars is appreciable, the estimated distortion of the CMB due to the increase in temperature $T$ cannot be detected with the current detector sensitivity. It is possible that further improvements in  estimated angular resolutions as well as in the precision of the temperature fluctuation measurements and experimental facilities such as the Large Hadron Collider (LHC) will make such effects as well as those of the photon anomalous magnetic moment observable. 

\par There has been a surge of interest to investigate quantum nonlinearity in state of the art optical experimental setups \citep{mar06}. The QED vacuum in an external field will reveal further interesting insights into processes such as electro-gravitational conversion \citep{pap77}. It will illuminate our further understanding of Lorentz Symmetry Breaking (LSB) in nonlinear electrodynamics \citep{cha12}. Some of the strongest magnetic fields in the universe are expected to exist around magnetars  \citep{ola14,bas08,kas14}. 
A strong magnetic field exists around the center of the galaxy \citep{eat13}. These objects with such strong magnetic fields, although contained in regions small relative to the cosmos, can still provide us with possibilities of observing nonlinear effects such as birefringence that can provide a handle to estimate physical quantities such as the photon anomalous magnetic moment and Faraday rotation \citep{eat13}. Proposals have been given to search for birefringence with the use of the time varying electromagnetic fields and high precision interferometry  \citep{gro15,zav09}.
	
	As long as the spatial and time inhomogeneitics are much larger than the Compton wavelength, the constant field approximation results will be reasonably accurate. More refined experimental observations of vacuum birefringence may facilitate a measurement of the photon anomalous magnetic moment. The BMV experiment \citep{cad14} is working on the vacuum birefringence measurements. The PVLAS experiment, which has been working for over two decades and proved that this extremely difficult measurement is feasible \citep{zav08,bre08,can08}, continues to make progress each year. This suggests that the measurements of the photon anomalous magnetic moment, even if indirect, may not be far away due to its close connection with the birefringence coefficients. The photon anomalous magnetic moment can be measured for low frequencies in view of the upcoming upscale experimental facilities for operation. Magnetars should provide an avenue for measurement through astroparticle physics in the large frequency limit. 

	Observational manifestations of nonlinear effects are feasible. Earlier works \citep{hey05,wan09} claim that QED nonlinear effects are detectable. Efforts to build an X-ray polarimeter are on the way. \cite{soff13} show the influence of magnetic vacuum birefringence on the polarization of magnetic neutron stars. A direct measurement of BMV would be a striking experimental proof of the fact that the nonlinearity in the vacuum is a reality for strong macroscopic electromagnetic fields. An appreciable signal of the Faraday rotation angle $\chi$ for the magnetized vacuum would be a new signature of the fundamental physics.

\section{Acknowledgements}

This paper is dedicated to the memory of Dr. Julian Mielniczuk, who unfortunately passed away in July 2016. We regret that he could not live to see this publication.

We would like to thank Professors Ken Roberts,  Victoria Kaspi, Guido Zavattini, Federico Della Valle, William Baylis, Gert Brodin, Rob Mann, Gerry McKeon and Janusz Sokol for their valuable suggestions to the revision of this draft. We also thank Niels Oppermann and Ue-Li Pen at CITA for directing us to the references on Faraday Rotation. S.R.Valluri would like to thank King's University College for their generous support in his research endeavors.

%,Niels Oppermann,Ue-Li Pen and Gerry McKeon 

\appendix

\section*{Appendix A}
\begin{flushleft}We will prove that \end{flushleft}
\begin{equation}\label{A1}
\frac{d}{dB}n_{\perp}(B)\geq0
\end{equation}
and subsequently
\begin{equation}\label{A2}
\frac{d}{dB} \langle H(B)\rangle \leq 0
\end{equation}
for the perpendicular mode.
We use 
\begin{equation}
n_{\perp}(B) =1+ \frac{1}{2}B^2 \frac{\partial^2 \mathcal{L}}{\partial\mathcal{G}^2}|^{\mathcal{F}=\frac{1}{2}B^2}_{\mathcal{G}=0}
\end{equation}

\begin{flushleft}and equation 63 of \cite{sha11} to get,\end{flushleft}

\begin{equation}\label{A4}
B^2 \frac{\partial^2 \mathcal{L}}{\partial\mathcal{G}^2 |^{\mathcal{F}=\frac{1}{2}B^2}_{\mathcal{G}=0}} = \frac{\alpha}{3\pi}B^2\frac{1}{2\mathcal{F}}\int_0^{\infty}\frac{dt}{t}\exp \left(\frac{-t}{b}\right)
\times\left[\frac{-3\coth t}{2t}+\frac{3}{2\sinh^2t}+t \coth t \right]
\end{equation} 
where $b= \frac{B}{B_{\rm cr}}$. Differentiating the RHS of equation \ref{A4} with respect to B we get
\begin{equation}
\frac{\alpha}{3\pi}\int^{\infty}_{0} \left(dt \frac{1}{b^2}\right)\exp\left(\frac{-t}{b}\right) \times\left[\frac{-3\coth t}{2t}+\frac{3}{2\sinh^2t}+t \coth t \right]
\end{equation}
Noting that 
\begin{equation}
\left[\frac{-3\coth t}{2t}+\frac{3}{2\sinh^2t}+t \coth t \right] \geq 0
\end{equation}
for each $t>0$ proves equation \ref{A1} and subsequently equation \ref{A2}. 
The derivation of equation \ref{2.1} will be provided in appendix B equation \ref{B23}. Equation \ref{2.4} provides the positive value of RHS for $0\leq B \leq \frac{1}{2} B_{\rm cr}$. 
For comparison the anomalous magnetic moment of an electron is \citep{cha12}
\begin{equation}
\mu_{\rm e,anom}=\frac{\alpha}{2\pi}\frac{1}{2}\frac{e}{m}.
\end{equation}

\begin{flushleft}so if we use \end{flushleft}
\begin{equation}
\mu_\gamma(B)\simeq  \frac{\alpha}{4\pi}\frac{2}{3}\frac{e}{m} = \frac{\alpha}{2\pi}\frac{1}{2}\frac{e}{m} \frac{2}{3}.
\end{equation}
$\mu_\gamma(B)$ is an $2/3$ order of magnitude of the anomalous magnetic moment of an electron $e^{-}$,where $e^{-}\cong 0.00115965 $  i.e.,
\begin{equation}\label{A9}
\mu_\gamma(B) \simeq \frac{2}{3} \mu_{\rm anom,{e^{-}}}
\end{equation}
Equation \ref{A9} provides an experimental upper bound for the photon in terms of the Bohr magneton
\citep{alt08} provides an experimental upper bound for $\mu_\gamma$ in terms of the Bohr magneton $\frac{1}{2}\frac{e}{m}$
\begin{equation}
\mu_\gamma(B)\sim {7.7 \times 10^{-4}} \mu_{\rm Bohr}
\end{equation}
\begin{equation}
\frac{d \langle H \rangle}{dB} \leq 0
 \end{equation}
implies
\begin{equation}
2 B \gamma_{\mathcal{G}\mathcal{G}}+ B^3 \gamma_{\mathcal{F}\mathcal{G}\mathcal{G}}\geq 0
\end{equation}
where 
\begin{equation}
 \gamma_{\mathcal{G}\mathcal{G}} =\frac{\partial^2 \mathcal{L}}{\partial\mathcal{G}^2}|^{\mathcal{F}=\frac{1}{2}B^2}_{\mathcal{G}=0}
\end{equation}
Where $\mathcal{L}$ is the Heisenberg-Euler Lagrangian and 

\begin{equation}
 \gamma_{\mathcal{F}\mathcal{G}\mathcal{G}} =\frac{\partial^3 \mathcal{L}}{\partial \mathcal{F}\partial\mathcal{G}^2}|^{\mathcal{F}=\frac{1}{2}B^2}_{\mathcal{G}=0}
\end{equation}

\begin{equation}
\frac{d^2 \langle H \rangle}{dB^2} \leq 0
 \end{equation}
implies
\begin{equation}
2\gamma_{\mathcal{G}\mathcal{G}}  + 5B^2 \gamma_{\mathcal{F}\mathcal{G}\mathcal{G}} + B^4  \gamma_{\mathcal{F} \mathcal{F}\mathcal{G}\mathcal{G}} \geq 0
\end{equation}
where 
\begin{equation}
 \gamma_{\mathcal{F}\mathcal{F}\mathcal{G}\mathcal{G}} =\frac{\partial^4 \mathcal{L}}{\partial \mathcal{F}\partial\mathcal{F}\partial\mathcal{G} \partial\mathcal{G}}|^{\mathcal{F}=\frac{1}{2}B^2}_{\mathcal{G}=0}
\end{equation}

\section*{Appendix B}

\begin{flushleft} We will prove equation \ref{2.4}. The starting point is equation \ref{12} of \cite{kar15} \end{flushleft}
\begin{equation}
\frac{\partial^2\mathcal{L}}{\partial \mathcal{G}^2|_{\mathcal{G}=0}} = \frac{1}{2\mathcal{F}}\frac{\alpha}{\pi}\left\lbrace 4 \zeta^{\prime}(-1,\chi)-\chi[2\zeta^{\prime}(0,\chi)-\ln(\chi)+\chi]-\frac{1}{6}(2\psi(\chi))+\chi^{-1}+1\right\rbrace
\end{equation}
with 
\begin{eqnarray}
\chi & = & \frac{m^2}{2\sqrt{2|\mathcal{F}|}} \times \left\lbrace
\begin{array}{cc}
1 & \rm{for}  \mathcal{F} \geq 0\\
i  & \rm{for} \mathcal{F} \leq 0
\end{array}\right\rbrace \\ \nonumber
(\hat a)_\pm & = & a_\mu\sigma^\mu_\pm 
\end{eqnarray}
\begin{equation}
\zeta^{\prime} (s,\chi)= \partial_s \zeta(s,\chi),
\end{equation}
where $\zeta(s,\chi)$ is the Hurwitz zeta function.

\begin{flushleft}In our case \end{flushleft}
\begin{equation}
\mathcal{F}= \frac{1}{2} B^{2}\geq 0
\end{equation}
For the case  $B>0$, use of the relation
\begin{equation}
\frac{d}{dB} \zeta^{\prime}(-1,1/(2B)) = \frac{1}{2}\frac{\left(-\ln \Gamma\left(\frac{1}{2B}\right))+\frac{1}{2}\ln (2\pi)-\frac{1}{2B}+\frac{1}{2}\right)}{B^2}\,\,\, ,
\end{equation}
gives:
\begin{equation}\label{B23}
\frac{d}{dB}\left(\frac{1}{2}B^2\gamma_{\mathcal{G}\mathcal{G}}\right)=\frac{\alpha}{4\pi}\left\lbrace\frac{2}{3}+\frac{1}{B^3}\left[ \frac{B}{3}\psi^{\prime}\left(1+\frac{1}{2B}\right)+\psi\left(\frac{1}{2B}\right) -2B\ln\Gamma\left(\frac{1}{2B}\right)+B \ln(2\pi)+ \ln(2B)       \right]\right\rbrace.
\end{equation}
Also, from the relation
\begin{equation}
\mu_{B}= - \frac{d}{dB} \langle H(B)\rangle
\end{equation}
we obtain equation \ref{2.1}.
We use the following inequality 
\begin{equation}
\psi^{\prime\prime}(1+h)\leq - \frac{1}{h^2}+\frac{1}{h^3}-\frac{1}{2}\frac{1}{h^4}+\frac{1}{6}\frac{1}{h^6} , h> 0
\end{equation}
plus similar inequalities for $\psi^{\prime}$,$\psi$ and $\ln \Gamma$.
We have for $h=\frac{1}{2B}$

\begin{equation}
 -\frac{1}{6} \frac{1}{B^4}\psi^{\prime \prime} (1+h)\geq \left(-\frac{1}{6}\frac{1}{B^4}          \right) \left(-\frac{1}{h^2} + \frac{1}{h^3} -\frac{1}{2}\frac{1}{h^4} +\frac{1}{6} \frac{1}{h^6}   \right).
\end{equation}

\section*{Appendix 3}
\begin{flushleft}We note that\end{flushleft}
\begin{equation}\label{C1}
\frac{d^2}{dB^2}n_{\perp}(B)\geq0
\end{equation}
and subsequently
\begin{equation}\label{C2}
\frac{d^2}{dB^2} \langle H(B)\rangle \leq 0
\end{equation}
We will use

\begin{eqnarray}
\frac{d^2}{dB^2}(\frac{1}{2}B^2_{\gamma\mathcal{G}\mathcal{G}}) =\frac{d}{dB} \frac{\alpha}{4\pi}\frac{2}{3}+\frac{1}{B^3}[\frac{B}{3}\psi^{\prime}(1+\frac{1}{2B})+\psi(\frac{1}{2B})-2B\ln\Gamma(\frac{1}{2B})+B\ln(2\pi)+\ln(2B)] , 0.44 \geq B \geq 0
\end{eqnarray}

\begin{flushleft}we will arrive at\end{flushleft}
\begin{equation}\label{C4}
\frac{d \mu(B)}{dB} \geq \frac{\alpha}{4\pi} \left( \frac{28}{45}  - \frac{156}{49} B^2 \right)\left(\frac{|\textbf{k}|}{m}\right) \sin ^2 \Theta .
\end{equation}
 
\begin{flushleft} For $0\leq0.44 B_{\rm cr}$ \end{flushleft}
\begin{equation}
\frac{\alpha}{4\pi} \left( \frac{28}{45}  - \frac{156}{49} B^2        \right) >0
\end{equation}
 because 
 \begin{equation}
 \frac{d^2}{dB^2} (B) = -\frac{d \mu(B)}{db} 
 \end{equation}
 we see that equation \ref{C2} is true in view.
 Similarly 
 \begin{equation}
 \frac{d^2}{dB^2} n_{\perp}(B) = \frac{d^2}{dB^2} (B) = -\frac{d \mu(B)}{dB} .
 \end{equation}

 \newpage

\newpage

\bibliography{Test1}
\bibliographystyle{aasjournal}
\end{document}